\documentclass[twocolumn,amsmath,amssymb,aps,prd,nofootinbib]{revtex4-1}
\usepackage{graphicx}
\usepackage{color}
\usepackage{soul}

\begin{document}

\title{Unified description of dark energy and dark matter within the generalized hybrid metric-Palatini theory of gravity}

\author{Paulo M. S\'a}

\email{pmsa@ualg.pt}

\affiliation{Departamento de F\'{\i}sica, Faculdade de Ci\^encias e Tecnologia, Universidade do Algarve, Campus de Gambelas, 8005-139 Faro, Portugal}

\date{June 5, 2020}

\begin{abstract}
The generalized hybrid metric-Palatini theory of gravity admits a scalar-tensor representation in terms of two interacting scalar fields. We show that, upon an appropriate choice of the interaction potential, one of the scalar fields behaves like dark energy, inducing a late-time accelerated expansion of the universe, while the other scalar field behaves like pressureless dark matter that, together with ordinary baryonic matter, dominates the intermediate phases of cosmic evolution. This unified description of dark energy and dark matter gives rise to viable cosmological solutions, which reproduce the main features of the evolution of the universe.
\end{abstract}

\maketitle

\section{Introduction}

General Relativity, proposed by Albert Einstein more than one hundred years
ago \cite{einstein-1916}, has been extremely successful in describing
gravitational phenomena on a vast range of scales --- from the Solar System to
the Universe itself ---, surpassing, to date, the most demanding experimental
tests \cite{will-2016}, including the recent detection of elusive
gravitational waves \cite{abbott-2016}.

However, notwithstanding this success, theories of gravity going beyond
General Relativity have been receiving increased attention for a variety of
reasons, among which the search for an alternative explanation to the observed
late-time accelerated expansion of the Universe stands out
\cite{riess-1998,perlmutter-1999}. The standard --- and simplest ---
explanation for this accelerated expansion, provided by the concordance
($\Lambda$CDM) model, resorts to a cosmological constant, but it turns out
that the theoretical value of this constant, resulting from quantum field
theory calculations, is overwhelmingly different from the one required by
observations. This is the well known cosmological-constant problem
\cite{weinberg-1989,martin-2012}, which, hopefully, could be avoided within
the framework of a modified theory of gravity.

A plethora of alternative theories to Einstein's General Relativity has been
considered in the literature (for a comprehensive review on the subject, see,
for instance, Ref.~\cite{berti-2015,nojiri-2017,heisenberg-2019}). In this
article, we will focus on a specific class of theories, namely, the hybrid
metric-Palatini theory of gravity \cite{harko-2012} and its generalized
version \cite{flanagan-2004,tamanini-2013}, which can be seen as extensions of
$f(R)$-gravity \cite{sotiriou-2010}.

In the hybrid metric-Palatini theory of gravity, the usual Einstein-Hilbert
action is supplemented with an additional term, $f(\mathcal{R})$, where
$\mathcal{R}$ is the Palatini curvature scalar, defined in terms of a
metric-independent connection, and $f$ is an arbitrary function
\cite{harko-2012}. This novel approach, which combines elements from both the
metric and Palatini formalisms, gives rise not only to a viable theory of
gravity, but also avoids some drawbacks of the original $f(R)$-gravity
\cite{capozziello-2015,koivisto-2013}.

The hybrid metric-Palatini theory of gravity admits a dynamically equivalent
scalar-tensor representation in terms of a long-range scalar field which plays
an active role in cosmology, as well as in galactic dynamics, without
conflicting with local experiments even if the scalar field is very light
\cite{harko-2012,capozziello-2013,capozziello-2013-b}. Other issues of
cosmological \cite{boehmer-2013,lima-2014,lima-2016,carloni-2015} and
astrophysical \cite{capozziello-2013-c,borka-2016,danila-2017,danila-2019}
relevance have also been addressed within this hybrid theory of gravity.
In particular, it has been shown that, for an appropriate choice
of the potential of the scalar field, solutions accounting for the late-time
acceleration of the universe can be obtained within this hybrid theory
\cite{harko-2012}, a result consolidated soon afterward by several other works
\cite{capozziello-2013,lima-2014,lima-2016,carloni-2015}.

A generalization of the hybrid metric-Palatini theory of gravity can be
achieved by considering the action to be an arbitrary function of {\em both}
the Ricci and Palatini scalars $f(R,\mathcal{R})$
\cite{flanagan-2004,tamanini-2013}. It turns out that this theory is
dynamically equivalent to a gravitational theory not with one, but two scalar
fields \cite{tamanini-2013}. The weak-field limit \cite{bombacigno-2019} and
black hole solutions \cite{rosa-2020} have been investigated within the
framework of this theory.

Cosmological solutions of this so-called generalized hybrid metric-Palatini
theory of gravity, including those showing periods of accelerated expansion of
the universe, were studied in the scalar-tensor representation (Einstein frame
\cite{tamanini-2013} and Jordan frame \cite{rosa-2017}), as well as in the
geometric representation \cite{rosa-2019}.

The purpose of the present article is to extend these analyses by
proposing a unified description of dark matter and dark energy.
To this end, we take advantage of the fact that the generalized hybrid
metric-Palatini theory of gravity admits a tensor representation with two
scalar fields. Then, one can choose a potential for these two
scalar fields such that one of them
plays the role of dark matter and, together
with ordinary baryonic matter, guarantees the existence of a long-lasting
matter-dominated era, and the other scalar field plays the role of dark
energy, inducing a late-time era of accelerated expansion of the universe.

Let us note that unified descriptions of dark matter and dark energy have also
been proposed in other contexts, such as, for instance, quintessence
\cite{sahni-2000}, generalized Chaplygin gas \cite{bento-2002,zhang-2006},
phantom cosmology \cite{capozziello-2006}, string landscape
\cite{liddle-2006,liddle-2008}, Salam-Sezgin six-dimensional supergravity
\cite{anchordoqui-2007,henriques-2009}, $k$-essence \cite{bose-2009},
generalized Galileon theories \cite{koutsoumbas-2018}, or mimetic gravity
\cite{dutta-2018}.

This article is organized as follows. In the next section we briefly present
the generalized hybrid metric-Palatini theory of gravity. In
Sect.~\ref{unified description}, choosing an appropriate potential for the
interacting scalar fields, we derive a simplified set of equations that
provide a unified description of dark matter and dark energy. Numerical
solutions of these equations are presented in Sect.~\ref{solutions}. Finally,
in Sect.~\ref{Conclusions}, we present our conclusions.

\section{The generalized hybrid metric-Palatini theory of
gravity\label{GHMP gravity}}

For the generalized hybrid metric-Palatini theory of gravity the action is
given by \cite{flanagan-2004,tamanini-2013}
\begin{align}
 S = \frac{1}{2\kappa^2} \int d^4x \sqrt{-g} f(R,\mathcal{R}),
    \label{action 1}
\end{align}
where $g$ is the determinant of the metric $g_{\alpha\beta}$, $R$ is the usual
Ricci scalar, defined in terms of the metric and the Christoffel connection
$\Gamma^{\mu}_{\alpha\beta}=\frac12 g^{\mu\nu} ( \partial_\alpha g_{\beta\nu}
+ \partial_\beta g_{\alpha\nu} - \partial_\nu g_{\alpha\beta})$, $\mathcal{R}$
is the Palatini scalar, constructed with the  metric and an independent
connection $\hat{\Gamma}^{\mu}_{\alpha\beta}$, such that $\mathcal{R} =
g^{\alpha\beta} ( \partial_\mu \hat{\Gamma}^{\mu}_{\alpha\beta} -
\partial_\beta \hat{\Gamma}^{\mu}_{\alpha\mu} + \hat{\Gamma}^{\mu}_{\nu\mu}
\hat{\Gamma}^{\nu}_{\alpha\beta} - \hat{\Gamma}^{\mu}_{\nu\beta}
\hat{\Gamma}^{\nu}_{\alpha\mu})$, and $f$ is an arbitrary (but smooth enough)
function of the Ricci and Palatini scalars.
Matter couples to the metric, but not to the independent connection
(for a review on modified theories of gravity in the Palatini approach
see Ref.~\cite{olmo-2011}).
The natural system of units will
be adopted throughout this article, implying  that
$\kappa=\sqrt{8\pi}/m_\texttt{P}$, where $m_\texttt{P} = 1/\sqrt{G} = 1.22
\times 10^{19} {\rm GeV}$ is the Planck mass.

As shown in Ref.~\cite{tamanini-2013}, action (\ref{action 1}) can be written
in the dynamically equivalent form
\begin{align}
 S = \frac{1}{2\kappa^2} \int d^4x \sqrt{-g}
     \left[ \phi R  - \frac{3}{2\xi}(\partial \xi)^2 - V(\phi,\xi) \right],
    \label{action 3}
\end{align}
where the scalars field $\phi$ and $\xi$ and the potential $V(\phi,\xi)$ are
related to function $f$ appearing in action (\ref{action 1}).

Note that this scalar-tensor representation differs significantly from the one
admitted by the original hybrid metric-Palatini theory of gravity (for a
derivation of the dynamically equivalent scalar-tensor representation of the
original theory see Ref.~\cite{harko-2012}). In particular, it contains an
extra scalar field, $\xi$, which, as will be seen below, turns out to be
crucial in our proposal for a unified description of dark energy and dark
matter.
Without going into details, we just mention that in the
original theory the scalar field is defined as
$\phi=df(\mathcal{R})/d\mathcal{R}$ \cite{harko-2012}, while in the
generalized version the two scalar fields are defined as $\xi=-\partial
f(R,\mathcal{R})/\partial \mathcal{R}$ and $\phi=-\xi+\partial
f(R,\mathcal{R})/\partial R$ \cite{tamanini-2013}.

Performing a conformal transformation $\tilde{g}_{\alpha\beta} =
\phi\hspace{0.5mm} g_{\alpha\beta}$ and redefining the scalar fields as $\phi=
\exp(\sqrt{2/3}\kappa \tilde{\phi})$ and $\xi=\kappa^2\tilde{\xi}^2/8$, one
obtains the above action in the Einstein frame, namely,
\begin{align}
 S = {}& {} \int d^4x \sqrt{-\tilde{g}} \bigg[ \frac{1}{2\kappa^2} \tilde{R}
   - \frac12 (\tilde{\nabla} \tilde{\phi})^2 \nonumber
 \\
  & - \frac12 e^{-\sqrt{2/3}\kappa \tilde{\phi}} (\tilde{\nabla}\tilde{\xi})^2
   - \tilde{V}(\tilde{\phi},\tilde{\xi}) \bigg],
        \label{action EF}
\end{align}
where the potential is defined as
\begin{align}
  \tilde{V}(\tilde{\phi},\tilde{\xi}) = \frac{1}{2\kappa^2}
  e^{-\sqrt{8/3}\kappa \tilde{\phi}}
  V (e^{\sqrt{2/3}\kappa \tilde{\phi}},\kappa^2 \tilde{\xi}^2/8).
    \label{generic potential}
\end{align}

Such actions, with two interacting scalar fields and non-standard kinetic
terms, motivated by alternative theories to Einstein's General Relativity, as
well as by supergravity and string theories, have received much attention in
the context of two-field inflationary models
\cite{berkin-1991,starobinsky-2001,dimarco-2003,lalak-2007,chakravarty-2016}.
In this article, we will be interested in later cosmological epochs, well
after the inflationary era, proposing a unified description of dark matter and
dark energy based on action (\ref{action EF}), in which the scalar fields
$\tilde{\xi}$ and $\tilde{\phi}$ play the role of dark matter and dark energy,
respectively.

In what follows, to avoid overloading the notation, the tildes in action
(\ref{action EF}) will be dropped.

\section{Unified description of dark energy and dark
matter\label{unified description}}

Let us now add radiation and ordinary baryonic matter to action~(\ref{action EF}),
both described by a perfect fluid with energy-momentum tensor
\begin{align}
 T^{\alpha\beta} = p g^{\alpha\beta}+(\rho+p) u^\alpha u^\beta,
\end{align}
where $u^{\alpha}$ denotes the four-velocity of an observer comoving with the
fluid; $\rho$ and $p$ are the energy density and pressure of the fluid,
respectively, related through the equation-of-state parameter $w=p/\rho$,
which takes the value $w_\texttt{R}=1/3$ for radiation and $w_\texttt{BM}=0$
for baryonic matter.

The Einstein equations are then
\begin{align}
 & R_{\alpha\beta} - \frac12 g_{\alpha\beta} R = 
 \kappa^2 \bigg\{ \nabla_\alpha \phi \nabla_\beta \phi
 -\frac12 g_{\alpha\beta}(\nabla\phi)^2 \nonumber
\\
 & \hspace{10mm} + \bigg[ \nabla_\alpha \xi \nabla_\beta \xi
 -\frac12 g_{\alpha\beta}(\nabla\xi)^2 \bigg] e^{-\sqrt{2/3}\kappa\phi} 
 - g_{\alpha\beta} V \nonumber
\\
 &  \hspace{10mm} + \bigg( \rho_\texttt{BM} + \frac43 \rho_\texttt{R} \bigg) u_\alpha u_\beta
 + \frac{\rho_\texttt{R}}{3} g_{\alpha\beta} \bigg\},
\end{align}
while the equations for the scalar fields $\phi$ and $\xi$ are given by
\begin{align}
 \nabla_\alpha \nabla^\alpha \phi
 + \frac{\kappa}{\sqrt6} (\nabla\xi)^2 e^{-\sqrt{2/3}\kappa\phi}
 - \frac{\partial V}{\partial \phi} = {}& 0,
\\
 \nabla_\alpha \nabla^\alpha \xi
 - \sqrt{\frac23} \kappa \nabla_\alpha\phi \nabla^\alpha \xi
 - \frac{\partial V}{\partial \xi} e^{\sqrt{2/3}\kappa\phi} = {}& 0.
\end{align}

Since current cosmological measurements constrain the present-time value of
the curvature density parameter $\Omega_k$ to be very small
\cite{ade-2016,aghanim-2018}, a spatially flat universe can be considered
without much loss of generality. Therefore, we assume a flat
Friedmann-Robertson-Walker metric,
\begin{align}
 ds^2 = -dt^2 + a^2(t) d\Sigma^2,
    \label{FRW metric}
\end{align}
where $a(t)$ is the scale factor and $d\Sigma^2$ is the metric of the
three-dimensional Euclidean space.

Taking into account that perfect-fluid energy-momentum conservation yields
$\rho_\texttt{R}\propto a^{-4}$ and $\rho_\texttt{BM}\propto a^{-3}$ for
radiation and baryonic matter, respectively, we obtain the equations for the
scale factor $a(t)$,
\begin{align}
 \left( \frac{\dot{a}}{a}\right)^2 = {}&
 \frac{\kappa^2}{3} \bigg[
 \frac{\dot{\phi}^2}{2}
 + \frac{\dot{\xi}^2}{2} e^{-\sqrt{2/3}\kappa \phi}
 + V(\phi,\xi) \nonumber
\\
 & + \rho_{\texttt{BM}0} \left( \frac{a_0}{a} \right)^3
 + \rho_{\texttt{R}0}
 \left( \frac{a_0}{a} \right)^4 \bigg],
   \label{Eq Friedman 1}
\\
 \frac{\ddot{a}}{a} = {}&
 -\frac{\kappa^2}{3} \bigg[
 \dot{\phi}^2
 + \dot{\xi}^2 e^{-\sqrt{2/3}\kappa\phi}
 - V(\phi,\xi) \nonumber
\\
 & + \frac12 \rho_{\texttt{BM}0} \left( \frac{a_0}{a} \right)^3
 + \rho_{\texttt{R}0} \left( \frac{a_0}{a} \right)^4 \bigg],
   \label{Eq Dotdota 1}
\end{align}
and for the scalar fields $\phi(t)$ and $\xi(t)$,
\begin{align}
 & \ddot{\phi}
 + 3\frac{\dot{a}}{a}\dot{\phi}
 + \frac{\kappa}{\sqrt{6}} \dot{\xi}^2 e^{-\sqrt{2/3}\kappa\phi}
 + \frac{\partial V}{\partial\phi}=0,
    \label{Eq phi 1}
\\
 & \ddot{\xi}
 + 3\frac{\dot{a}}{a}\dot{\xi}
 - \sqrt{\frac23}\kappa \dot{\phi}\dot{\xi}
 + \frac{\partial V}{\partial\xi} e^{\sqrt{2/3}\kappa\phi}=0,
    \label{Eq xi 1}
\end{align}
where an overdot denotes a derivative with respect to time $t$ and a subscript
$0$ denotes the  value of a variable at the present time. For the present-time
value of the energy densities of radiation and baryonic matter we take,
respectively, $\rho_{\texttt{R}0}=9.02\times10^{-128} m_\texttt{P}^4$ and
$\rho_{\texttt{BM}0}=8.19\times10^{-125} m_\texttt{P}^4$, having assumed
$H_0\equiv(\dot{a}/a)_0=67\,{\rm km}\,{\rm s}^{-1}\,{\rm Mpc}^{-1}$.

Now, in order to allow for an unified description of dark energy and dark
matter that is, at least in broad lines, in agreement with cosmological data,
one needs to choose the potential $V(\phi,\xi)$ appropriately. Taking into
account that it should be constructed with exponentials of $\phi$ and powers
of $\xi^2$ [see Eq.~(\ref{generic potential})], our choice is
\begin{align}
 \begin{split}
 V(\phi,\xi) & = V_a e^{-\frac{\lambda\kappa}{\sqrt6}\phi}
    \bigg[ 1+\bigg( \frac{\xi}{\xi_a} \bigg)^2  \bigg] \\
             & = V_a e^{-\frac{\lambda\kappa}{\sqrt6}\phi}
             + \frac12 m_\xi^2 (\phi) \xi^2,
 \end{split}
    \label{potential DE-DM}
\end{align}
where $V_a>0$, $\lambda$, and $\xi_a$ are constants, and $m_\xi$ denotes the
$\phi$-dependent mass of the scalar field $\xi$,
\begin{align}
 m_\xi(\phi) = \sqrt{\frac{2V_a}{\xi_a^2}} e^{-\frac{\lambda\kappa}{2\sqrt6}\phi}.
   \label{mass xi}
\end{align}

Before proceeding with the analysis of the above system of equations, let us
point out that in Ref.~\cite{tamanini-2013} other choices were made for the
potential (\ref{generic potential}). There, in one of the analyzed cases, the
potential was chosen to depend exclusively on the scalar field $\phi$ and to
be of the quintessential type, namely, $V\propto
\exp(-\lambda\kappa\phi/\sqrt6)$; this ensures a late-time accelerated
expansion of the universe, but at the same time requires dark matter to be
introduced ``by hand" in the standard matter sector. Another potential,
depending both on $\phi$ and $\xi$, was also considered, namely, $V \propto
\xi^\lambda \exp(-\lambda\kappa\phi/\sqrt6)$; in this case, the existence of a
global attractor corresponding to an accelerated solution (for
$\lambda<2\sqrt3$) was proven, but the universe does not undergo an
intermediate phase of matter domination.

In the present article, the analysis of Ref.~\cite{tamanini-2013}
is extended in order to provide a unified description of dark matter and dark energy.
This requires an appropriate choice of the interaction potential, namely, a choice
that guarantees that the
scalar field $\phi$ behaves like dark energy, dominating the dynamics of late-time
cosmological evolution and yielding an accelerated expansion of the universe, while
the scalar field $\xi$ behaves like pressureless dark matter that, together with
ordinary baryonic matter, dominates the intermediate phases of cosmic evolution.
As will be shown below, this can be accomplished by choosing a
potential of the form given by Eq.~(\ref{potential DE-DM}), with the appropriate 
values of the constants $V_a$, $\xi_a$, and $\lambda$.

The system of equations (\ref{Eq Friedman 1})--(\ref{Eq xi 1}) can be
considerably simplified by taking into account that, for $m_\xi \gg H$, the
scalar field $\xi$ oscillates rapidly around its minimum\footnote{This
condition for $m_\xi$, which translates into $\xi_a^2 \ll 2 V_a
(\dot{a}/a)^{-2} \exp \{ -\lambda\kappa\phi / \sqrt6 \}$, is satisfied as long
as the value of $\xi_a$ in the  potential (\ref{potential DE-DM}) is chosen to
be small enough.}, behaving like a nonrelativistic dark-matter fluid with
equation of state $\big<p_\texttt{DM}\big>=0$ \cite{turner-1983}, where the
brackets $\big< ... \big>$ denote the average over an oscillation.

We define the energy density and pressure of the scalar field $\xi$ as
\begin{align}
 \rho_\texttt{DM} = \frac{\dot{\xi}^2}{2} e^{-\sqrt{2/3}\kappa\phi}
 + \frac12 m_\xi^2(\phi) \xi^2,
\\
 p_\texttt{DM} = \frac{\dot{\xi}^2}{2} e^{-\sqrt{2/3}\kappa\phi}
 - \frac12 m_\xi^2(\phi) \xi^2.
\end{align}
This differs from the situation considered in Ref.~\cite{turner-1983} because
of the non-standard kinetic and mass terms, both depending explicitly on the
scalar field $\phi$. However, since this field changes very slowly in
comparison to $\xi$, the approach used in Ref.~\cite{turner-1983} can be
applied here.

Let us then multiply Eq.~(\ref{Eq xi 1}) by $\dot{\xi}$ and use the above
definition of $\rho_\texttt{DM}$ to obtain
\begin{align}
& \dot{\rho}_\texttt{DM}
 +3\frac{\dot{a}}{a}\dot{\xi^2} e^{-\sqrt{2/3}\kappa\phi}
 +\frac{(\lambda-2)\kappa}{2\sqrt6} m_\xi^2 \xi^2 \dot{\phi} \nonumber
 \\
& \hspace{15mm} + \sqrt{\frac23} \kappa \dot{\phi} \left( \rho_\texttt{DM} -
   \dot{\xi}^2 e^{-\sqrt{2/3}\kappa\phi} \right) = 0.
\end{align}

Averaging over an oscillation period and taking into account that
$\big<p_\texttt{DM}\big>=0$ implies $\big< \xi^2 \big>
=\rho_\texttt{DM}/m_\xi^2$ and $\big< \dot{\xi}^2 \big> = \rho_\texttt{DM}
e^{\sqrt{2/3}\kappa\phi}$, the above equation can be written as
\begin{align}
 \dot{\rho}_\texttt{DM}
 + 3\frac{\dot{a}}{a} \rho_\texttt{DM}
 + \frac{(\lambda-2)\kappa}{2\sqrt6} \rho_\texttt{DM} \dot{\phi} = 0,
\end{align}
which yields the solution
\begin{align}
 \rho_\texttt{DM} = C \left( \frac{a_0}{a} \right)^3
 e^{\frac{(2-\lambda)\kappa}{2\sqrt6}\phi},
   \label{rho dark matter}
\end{align}
where $C$ is an integration constant.

As expected, the energy density of dark matter depends on the scale factor as
$a^{-3}$.
This is a direct consequence of the fact that we have chosen
the potential~(\ref{potential DE-DM}) to be quadratic on the scalar field $\xi$;
oscillations around the minimum of the potential mimic a nonrelativistic
matter fluid with vanishing pressure.
But $\rho_\texttt{DM}$ also depends on the field $\phi$, through the
exponential factor, reflecting the existence of a direct coupling between the
two fields $\phi$ and $\xi$ in action (\ref{action EF})
and in the potential (\ref{potential DE-DM}). As will be shown
in Sect.~\ref{solutions}, such dependence has implications on the cosmic
evolution, namely, it leads to a non-simultaneous peaking of the energy densities
of dark matter and baryonic matter.
For a general discussion of a two-scalar-field model for the interaction of
dark energy and dark matter see Ref.~\cite{bertolami-2012}.

Now, using Eq.~(\ref{rho dark matter}), the system of Eqs.~(\ref{Eq Friedman
1})--(\ref{Eq xi 1}) can be considerably simplified, yielding:
\begin{align}
 \left( \frac{\dot{a}}{a}\right)^2 = {}&
 \frac{\kappa^2}{3} \bigg[
 \frac{\dot{\phi}^2}{2}
 + V_a e^{-\frac{\lambda\kappa}{\sqrt{6}}\phi}
 + \rho_{\texttt{R}0} \left( \frac{a_0}{a} \right)^4 \nonumber
\\
 & + \left( \rho_{\texttt{BM}0} + C e^{\frac{(2-\lambda)\kappa}{2\sqrt6}\phi} \right)
 \left( \frac{a_0}{a} \right)^3
  \bigg],
    \label{Eq Friedman 2}
\\
 \frac{\ddot{a}}{a} = {}&
 -\frac{\kappa^2}{3} \bigg[ \dot{\phi}^2
 - V_a e^{-\frac{\lambda\kappa}{\sqrt{6}}\phi}
 + \rho_{\texttt{R}0} \left( \frac{a_0}{a} \right)^4 \nonumber
\\
 & +  \frac12 \left( \rho_{\texttt{BM}0}
 + C e^{\frac{(2-\lambda)\kappa}{2\sqrt6}\phi} \right)
 \left( \frac{a_0}{a} \right)^3
  \bigg],
    \label{Eq Dotdota 2}
\end{align}
and
\begin{align}
 & \ddot{\phi} + 3\frac{\dot{a}}{a}\dot{\phi}
 + \frac{(2-\lambda)\kappa C}{2\sqrt6} \left( \frac{a_0}{a} \right)^3
   e^{\frac{(2-\lambda)\kappa}{2\sqrt6}\phi} \nonumber
\\
 & \hspace{40mm}
 - \frac{\lambda\kappa V_a}{\sqrt6} e^{-\frac{\lambda\kappa}{\sqrt6}\phi}=0.
    \label{Eq phi 2}
\end{align}

Instead of the comoving time $t$ let us use a new variable $u$ related to the
redshift $z$,
\begin{align}
  u = - \ln (1+z) = -\ln \left( \frac{a_0}{a} \right).
\end{align}

In terms of this new variable, Eq.~(\ref{Eq phi 2}) becomes
\begin{align}
 \phi_{uu} = {}&
 - \bigg\{ \bigg[ \frac{\ddot{a}}{a}
 + 2 \bigg( \frac{\dot{a}}{a} \bigg)^2 \bigg]\phi_u
 -\frac{\lambda\kappa V_a}{\sqrt6} e^{-\frac{\lambda\kappa}{\sqrt6}\phi} \nonumber
\\
 & +\frac{(2-\lambda)\kappa C}{2\sqrt6}
   e^{\frac{(2-\lambda)\kappa}{2\sqrt6}\phi} e^{-3u} \bigg\}
 \left( \frac{\dot{a}}{a} \right)^{-2},
   \label{Eq phi 3}
\end{align}
where the subscript $u$ denotes a derivative with respect to $u$;
$(\dot{a}/a)^2$ and $\ddot{a}/a$ are function of $u$, $\phi$, and $\phi_u$
given by
\begin{align}
 \left( \frac{\dot{a}}{a} \right)^2 = {}&
  2\kappa^2  \bigg[ V_a e^{-\frac{\lambda\kappa}{\sqrt6}\phi}
 +\left( \rho_{\texttt{BM}0}
 + C e^{\frac{(2-\lambda)\kappa}{2\sqrt6}\phi}\right) e^{-3u} \nonumber
\\
 & \hspace{-0.5mm} + \rho_{\texttt{R}0}e^{-4u} \bigg]
  \left( 6-\kappa^2 \phi_u^2 \right)^{-1},
   \label{Eq friedman 3}
\\
 \frac{\ddot{a}}{a} = {}&
 \frac{\kappa^2}{6} \bigg\{
  4\kappa^2 \bigg[ V_a e^{-\frac{\lambda\kappa}{\sqrt6}\phi}
  +\left( \rho_{\texttt{BM}0}
  + C e^{\frac{(2-\lambda)\kappa}{2\sqrt6}\phi}\right) e^{-3u} \nonumber
\\
  & \hspace{-0.5mm} + \rho_{\texttt{R}0}e^{-4u} \bigg]
   \phi_{u}^2 \left( \kappa^2 \phi_u^2 - 6 \right)^{-1}
 + 2V_a e^{-\frac{\lambda\kappa}{\sqrt6}\phi} \nonumber
\\
 & \hspace{-0.5mm} - \left( \rho_{\texttt{BM}0}
 + C e^{\frac{(2-\lambda)\kappa}{2\sqrt6}\phi}\right) e^{-3u}
 - 2\rho_{\texttt{R}0}e^{-4u} \bigg\}.
   \label{Eq Dotdota 3}
\end{align}

The above equations (\ref{Eq phi 3})--(\ref{Eq Dotdota 3}), which provide a
unified description of dark matter and dark energy within the generalized
hybrid metric-Palatini theory of gravity, will be solved numerically in the
next section.

For later convenience let us introduce the density parameters for radiation,
baryonic matter, dark matter, and dark energy, respectively,
\begin{align}
 & \Omega_\texttt{R} = \frac{\rho_\texttt{R}}{\rho_c}
 = \frac{\kappa^2}{3}\rho_{\texttt{R}0} e^{-4u}
 \left( \frac{\dot{a}}{a} \right)^{-2},
\\
 & \Omega_\texttt{BM} = \frac{\rho_\texttt{BM}}{\rho_c}
 = \frac{\kappa^2}{3}\rho_{\texttt{BM}0} e^{-3u}
 \left( \frac{\dot{a}}{a} \right)^{-2},
\\
 & \Omega_\texttt{DM} = \frac{\rho_\texttt{DM}}{\rho_c}
 = \frac{\kappa^2}{3} C e^{\frac{(2-\lambda)\kappa}{2\sqrt6}\phi}
 e^{-3u} \left( \frac{\dot{a}}{a} \right)^{-2},
\\
 & \Omega_\texttt{DE} = \frac{\rho_\texttt{DE}}{\rho_c}
 = \frac{\kappa^2}{3} \bigg[ \frac{\phi_u^2}{2}
 + V_a e^{-\frac{\lambda\kappa}{\sqrt6}\phi}
 \left( \frac{\dot{a}}{a} \right)^{-2} \bigg],
\end{align}
and the effective equation-of-state parameter,
\begin{align}
 w_{\rm eff} =
 \frac13 \bigg[ 1 - \Omega_\texttt{BM} - \Omega_\texttt{DM} - \Omega_\texttt{DE}
 \bigg( 1- 3 \frac{p_\texttt{DE}}{\rho_\texttt{DE}} \bigg) \bigg],
    \label{eos parameter}
\end{align}
where $\rho_c=(3/\kappa^2)(\dot{a}/a)^2$ is the critical density, and
$\rho_\texttt{DE}$ and $p_\texttt{DE}$ are the energy density and pressure of
dark energy, respectively, given by
\begin{align}
 \rho_\texttt{DE} = \frac12 \left( \frac{\dot{a}}{a} \right)^2 \phi_u^2
 + V_a e^{-\frac{\lambda \kappa}{\sqrt6}\phi},
\label{rhoDE}
 \\
 p_\texttt{DE} = \frac12 \left( \frac{\dot{a}}{a} \right)^2 \phi_u^2
 - V_a e^{-\frac{\lambda \kappa}{\sqrt6}\phi}.
\label{pDE}
\end{align}

\section{Numerical Solutions\label{solutions}}

Let us now solve Eqs.~(\ref{Eq phi 3})--(\ref{Eq Dotdota 3}) numerically.

Since we want to include in our numerical simulations the transition from a
radiation- to a matter-dominated era and, afterward, from a matter- to a
dark-energy-dominated one, we choose to start the integration of Eqs.~(\ref{Eq
phi 3})--(\ref{Eq Dotdota 3}) at $u_i=-30$, corresponding to redshift $z\simeq
10^{13}$ (well inside the radiation-dominated era of the evolution of the
universe).

Having verified that the numerical solutions are quite insensitive to the
initial values of the scalar field $\phi$ and its derivative, we choose in all
simulations $\phi(u_i)=10^{-3} m_\texttt{P}$ and $\phi_u({u_i})=10^{-5}
m_\texttt{P}$.

\begin{figure}[t]
\includegraphics[width=86mm]{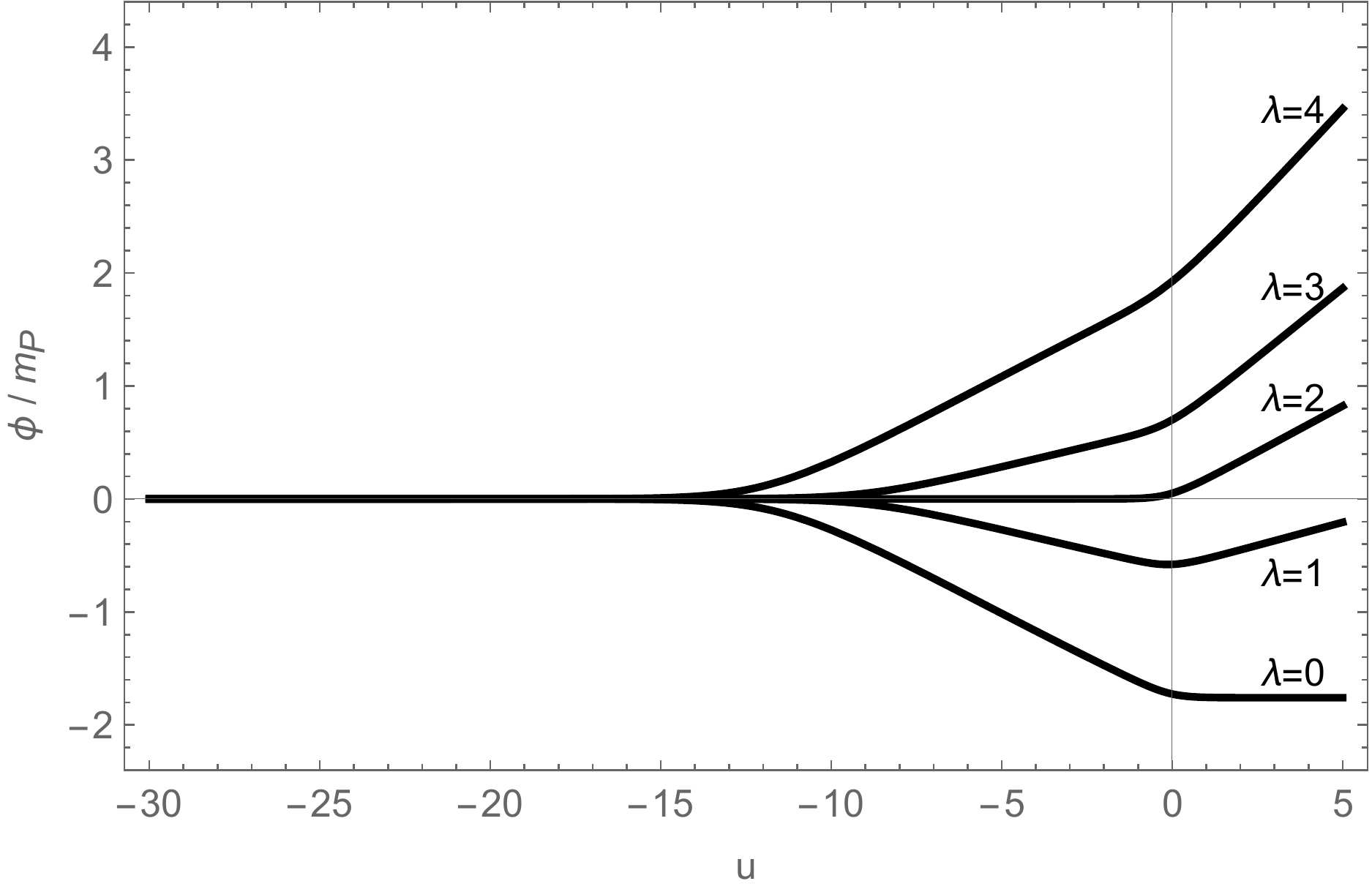}
\caption{Evolution of the scalar field $\phi$, which plays the role of dark
energy, for different values of the parameter $\lambda$. At a certain point,
this field becomes dominant and induces an accelerated phase of expansion of
the universe. Present time corresponds to $u_0=-\ln(1+z_0)=0$.}
 \label{Fig phi for different lambda}
\end{figure}

We want to guarantee that, at the present time $u_0=0$, the energy densities
of dark matter and dark energy, as well as radiation and ordinary baryonic
matter, are in agreement with current cosmological measurements
\cite{ade-2016,aghanim-2018}. Since these measurements indicate that dark
energy and dark matter contribute, at the present time, with about $69\%$ and
$26\%$, respectively, of the total energy density of the universe, we take
$\rho_{\texttt{DE}0}=1.13\times10^{-123} m_\texttt{P}^4$ and
$\rho_{\texttt{DM}0} = 4.25\times10^{-124} m_\texttt{P}^4$. We should then
impose that the constants $\lambda$, $V_a$, and $C$ satisfy the conditions
\begin{align}
 & \frac12 \bigg[ \left( \frac{\dot{a}}{a} \right)^2 \phi_{u}^2 \bigg]_{u=u_0}
  + V_a e^{-\frac{\lambda\kappa}{\sqrt6}\phi_0}  = \rho_{\texttt{DE}0},
     \label{condition 1}
\\
 & C e^{\frac{(2-\lambda)\kappa}{2\sqrt6}\phi_0} = \rho_{\texttt{DM}0}.
     \label{condition 2}
\end{align}
Since this can be achieved with just two of the above three constants, we
leave $\lambda$ as a free parameter and satisfy these conditions by choosing
appropriately $V_a$ and $C$. Therefore, to different values of the parameter
$\lambda$ will correspond different pictures of cosmic evolution, all of them
with the correct present-time values of the density parameters for radiation,
$\Omega_{\texttt{R}0}$, baryonic matter, $\Omega_{\texttt{BM}0}$, dark matter,
$\Omega_{\texttt{DM}0}$, and dark energy, $\Omega_{\texttt{DE}0}$, but with
different dynamics in the past. Naturally, we will focus our attention on the
values of the parameter $\lambda$ for which these past dynamics are, at least
in broad lines, in agreement with cosmological observations.

Let us recall that the constant $\xi_a$, determining the mass
of the scalar field $\xi$ [see Eq.~(\ref{mass xi})], is constrained by the
condition $m_\xi \gg H$, required to guarantee that the scalar field $\xi$
oscillates rapidly around its minimum, thus mimicking a pressureless
dark-matter fluid. This condition can always be satisfied by choosing a small
enough value of $\xi_a$.

Our numerical simulations show that, initially, the scalar field $\phi$, which
plays the role  of dark energy, remains practically constant. Depending on the
value of the parameter $\lambda$, at a certain point this field starts
evolving, eventually becoming dominant and inducing an accelerated phase of
expansion of the universe (see Fig.~\ref{Fig phi for different lambda}).

This behavior of the field $\phi$ implies that, in the earlier phases of
evolution, the energy density of dark matter $\rho_{\texttt{DM}}$ decreases as
$a^{-3}$, exactly as ordinary baryonic matter; for later times, depending on
the value of $\lambda$, it decreases faster or slower than $a^{-3}$, the
exception being the case $\lambda=2$ for which the exponential in
Eq.~(\ref{rho dark matter}) is always equal to one.

The different eras of evolution of the universe --- radiation dominated,
matter dominated, and dark energy dominated --- are shown in Fig.~\ref{Fig
Omega for lambda 2} for the case $\lambda=2$, for which
$V_a=1.32\times10^{-123}m_\texttt{P}^4$ and
$C=4.25\times10^{-124}m_\texttt{P}^4$ are required in order to satisfy the
conditions given by Eqs.~(\ref{condition 1}) and (\ref{condition 2}).

\begin{figure}[t]
\includegraphics[width=86mm]{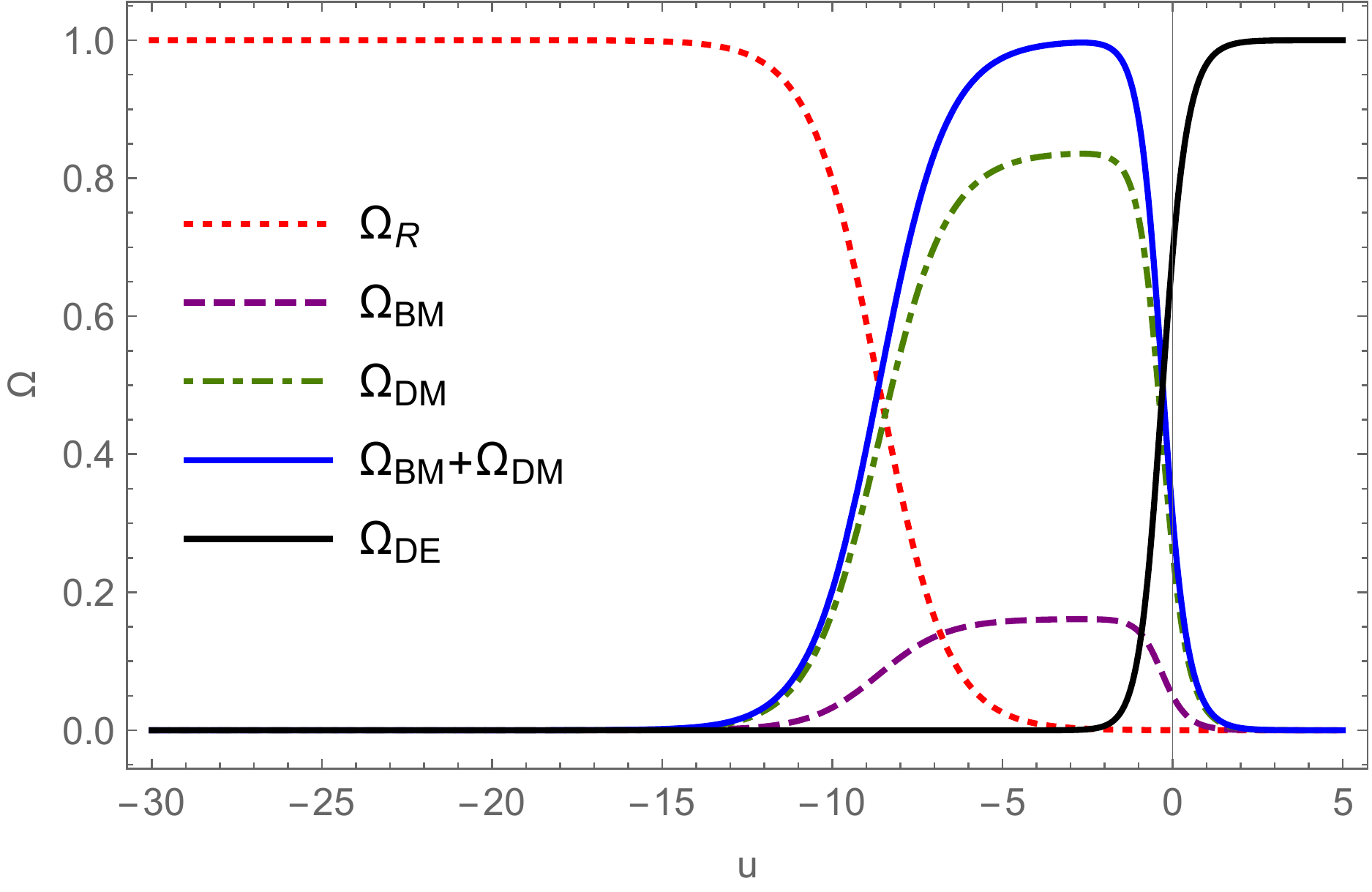}
\caption{Evolution of the density parameters of radiation, baryonic and dark
matter, and dark energy for $\lambda=2$. The transition from a radiation- to a
matter-dominated universe occurs well after the primordial nucleosynthesis
epoch; the (baryonic and dark) matter-dominated era lasts long enough for
structure formation to occur; the transition to a dark-energy-dominated
universe takes place in a recent past. At the present time $u_0=0$ the density
parameters are $\Omega_{\texttt{DE}0}=0.69$, $\Omega_{\texttt{DM}0}=0.26$,
$\Omega_{\texttt{BM}0}=0.05$, and $\Omega_{\texttt{R}0}=5\times10^{-5}$.}
 \label{Fig Omega for lambda 2}
\end{figure}

Initially, the dynamics of the universe is dominated by radiation and
$\Omega_\texttt{R}\simeq1$. Primordial nucleosynthesis, taking place at
redshift $z\simeq10^8$ ($u\simeq-18$), is well inside the radiation-dominated
era. At redshift $z\simeq10^6$ ($u\simeq-14$) begins the transition to a
matter-dominated universe and radiation-matter equality is achieved at
redshift $z\simeq 10^4$ ($u\simeq-9$). During this matter-dominated era, the
energy density of ordinary baryonic matter is a fraction of the energy density
of dark matter and $\Omega_\texttt{BM}+\Omega_\texttt{DM}\simeq1$. Finally, at
redshift $z\simeq 1$ ($u\simeq-0.7$), the scalar field $\phi$ becomes dominant
and the transition from a matter- to a dark-energy-dominated universe begins,
giving rise to accelerated expansion. At the present time ($u_0=0$), the
density parameters of dark energy, dark matter, baryonic matter, and radiation
are, respectively, $\Omega_{\texttt{DE}0}=0.69$, $\Omega_{\texttt{DM}0}=0.26$,
$\Omega_{\texttt{BM}0}=0.05$, and $\Omega_{\texttt{R}0}=5\times10^{-5}$.

The above ``history" of the universe reproduces, at least in broad lines, the
main features of the observed cosmic evolution. But, naturally, it depends on
the value of the parameter $\lambda$: the more $\lambda$ differs from 2, the
more this ``history" deviates from the standard one.

These deviations are mainly of three types (see Fig.~\ref{Fig Omega for lambda
0}, for the case $\lambda=0$; the case $\lambda=4$ is quite similar).

\begin{figure}[t]
\includegraphics[width=83.5mm]{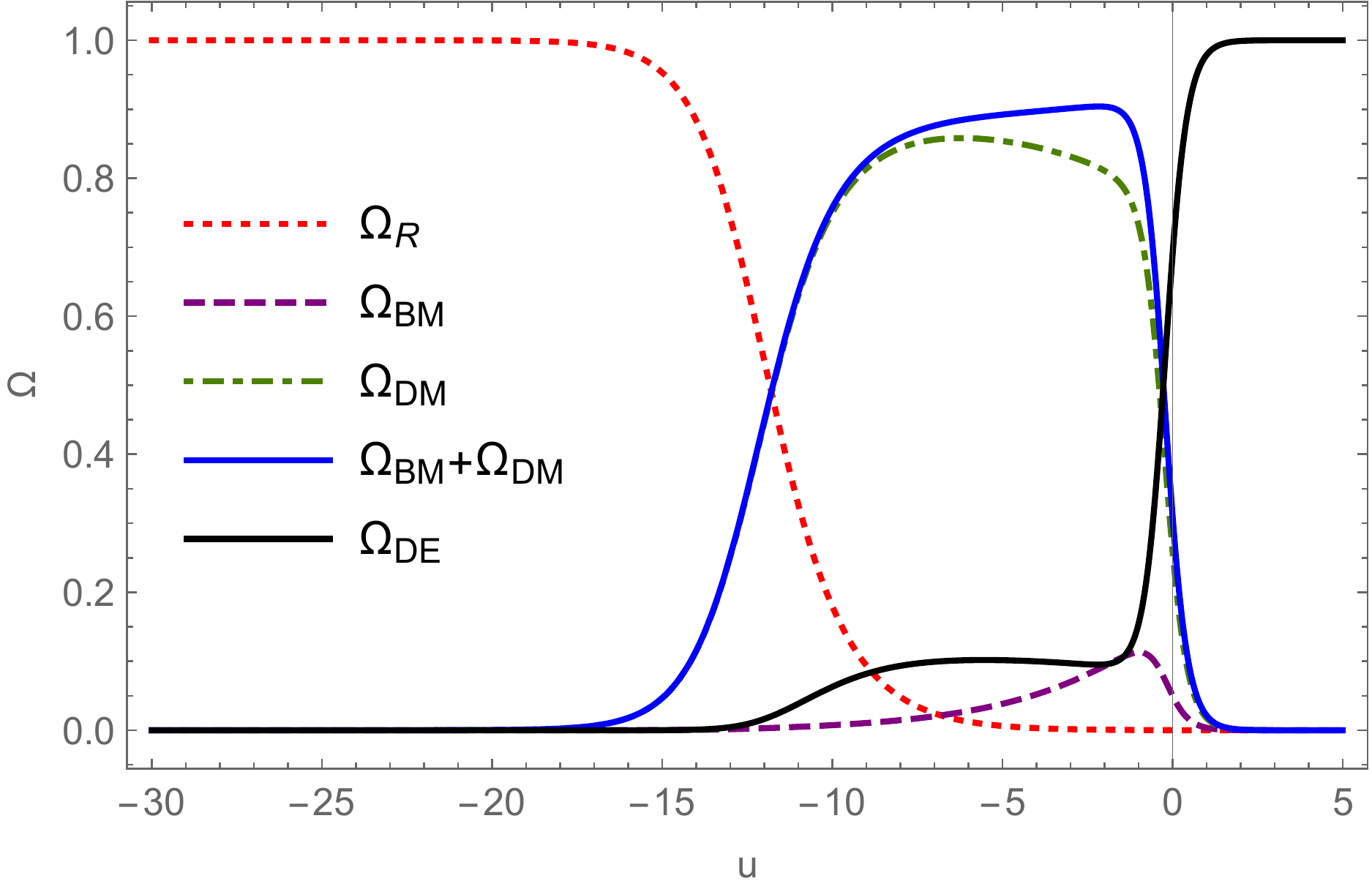}
\caption{Evolution of the density parameters of radiation, baryonic and dark
matter, and dark energy for $\lambda=0$ (similar to the case $\lambda=4$). The
transition to a matter-dominated universe occurs much earlier than in the case
$\lambda=2$, but still after the primordial nucleosynthesis epoch. Contrarily
to the case $\lambda=2$, the peaks of $\Omega_\texttt{BM}$ and
$\Omega_\texttt{DM}$ do not occur simultaneously. During the matter-dominated
era, the energy density of dark energy is already a noticeable fraction of the
total energy density. At the present time $u_0=0$ the density parameters are
$\Omega_{\texttt{DE}0}=0.69$, $\Omega_{\texttt{DM}0}=0.26$,
$\Omega_{\texttt{BM}0}=0.05$, and $\Omega_{\texttt{R}0}=5\times10^{-5}$.}
 \label{Fig Omega for lambda 0}
\end{figure}

First, the transition from a radiation- to a matter-dominated era occurs
earlier; our numerical simulations show that, for $|\lambda-2|\gtrsim2$, this
transition take place so early that it enters in conflict with primordial
nucleosynthesis. Since this is a conflict one wants to avoid, such values of
$\lambda$ should not be considered.

Second, as already mentioned above, for $\lambda\neq2$ the energy density of
dark matter, contrarily to ordinary baryonic matter, does not decrease exactly
as $a^{-3}$.
It also depends on the scalar field $\phi$ through an
exponential factor [see Eq.~(\ref{rho dark matter})],
which arises due to the existence of a direct coupling
between the two fields $\phi$ and $\xi$ in action (\ref{action EF})
and in the potential (\ref{potential DE-DM}). This implies
that the ratio between the energy densities of baryonic and dark matter is not
a constant anymore, but a quantity that depends on the dark energy field,
namely, $\rho_\texttt{BM}/\rho_\texttt{DM} \propto \exp \left[
\frac{(\lambda-2)\kappa}{2\sqrt6} \phi \right]$. Therefore, for $\lambda\neq2$
the peaks of $\Omega_\texttt{BM}$ and $\Omega_\texttt{DM}$ do not occur
simultaneously.

Third, the dark energy field $\phi$ becomes relevant at earlier stages of the
universe's evolution; our numerical simulations show that, for
$|\lambda-2|\gtrsim1$, the energy density of dark energy is already a
noticeable fraction of the total energy density during the matter-dominated
era. In the case shown in Fig.~\ref{Fig Omega for lambda 0}, at redshift
$z\simeq10^3$ ($u=-7$) the energy density of dark energy is already $1/9$ of
the combined energy densities of dark and baryonic matter.

\begin{figure}[t]
\includegraphics[width=86mm]{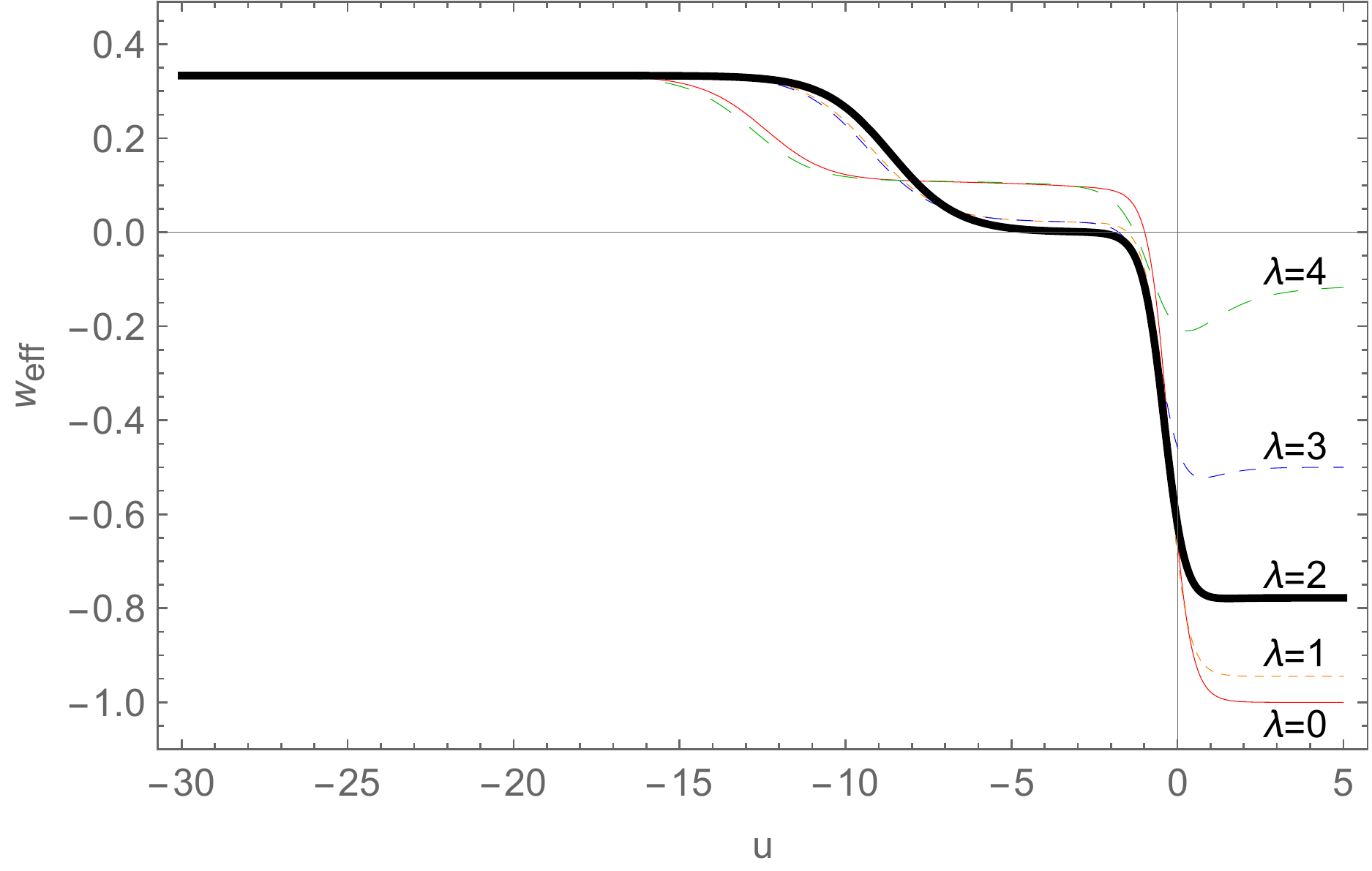} \caption{Evolution of
the effective equation-of-state parameter $w_{\rm eff}$ for different values
of $\lambda$. Initially, the evolution is dominated by radiation, implying
$w_{\rm eff}=1/3$ for all $\lambda$'s. At certain point, baryonic and dark
matter become dominant and $w_{\rm eff}$ decreases gradually from $1/3$ to a
near-zero value which depends on $\lambda$. In recent times, dark energy
becomes dominant and $w_{\rm eff}$ assumes negatives values; for $w_{\rm
eff}\leq -1/3$ it signals an accelerated expansion of the universe.}
 \label{Fig w for different lambda}
\end{figure}

Let us now analyze the evolution of the effective equation-of-state parameter
$w_{\rm eff}$ (see Fig.~\ref{Fig w for different lambda}), circumscribing
ourselves to values of the parameter $\lambda$  that correspond, at least
qualitatively, to the observed evolution of the universe, namely,
$0\lesssim\lambda\lesssim4$.

Initially, for all considered values of $\lambda$, the energy densities of
both baryonic and dark matter, as well as dark energy, are negligible, and the
dynamics of evolution is dominated by radiation, implying $w_{\rm eff}=1/3$.

Then, at a certain point, matter (baryonic and dark) becomes dominant and the
effective equation-of-state parameter decreases gradually from $1/3$ to zero
(a transition from a radiation- to a matter-dominated universe takes place).
However, for $|\lambda-2|\gtrsim1$, this behavior is changed because, during
the matter-dominated era, the energy density of dark energy is not negligible
anymore, contributing noticeably to $w_{\rm eff}$, which, therefore, takes
a value higher than zero (cases $\lambda=0$ and $\lambda=4$ in Fig.~\ref{Fig w
for different lambda}).

Finally, in recent times, the dark energy field $\phi$ becomes dominant and
the effective equation-of-state parameter assumes negative values.
Asymptotically, $w_{\rm eff}$ tends  to $-1+\lambda^2/18$. Indeed, for
negligible $\rho_\texttt{R}$, $\rho_\texttt{BM}$, and $\rho_\texttt{DM}$,
Eqs.~(\ref{Eq phi 3})--(\ref{Eq Dotdota 3}) yield the approximate solution
$\phi_u = \lambda/(\sqrt6 \kappa)$ and $V_a \exp ( -\lambda\kappa\phi/\sqrt6 )
= (3-\lambda^2/12)(H/\kappa)^2$, which, upon substitution in Eqs.~(\ref{eos
parameter})--(\ref{pDE}), gives $w_{\rm eff}=-1+\lambda^2/18$. This means that
for $\lambda\leq2\sqrt3$ the universe enters an everlasting period of
accelerated expansion. Note, however, that for values of $\lambda$ slightly
above $2\sqrt3$ the universe can experience a period of accelerated expansion,
but only temporary. For $\lambda=0$, the effective equation-of-state parameter
tends asymptotically to the value $-1$, which is not unexpected, since in this
case the potential given by Eq.~(\ref{potential DE-DM}) becomes simply
$V(\xi)=V_a+m_\xi^2 \xi^2/2$, corresponding to a cosmological constant plus a
(dark) matter field.

The above proposed unified description of dark matter and dark energy
depends, naturally, on the form of the potential~(\ref{potential DE-DM});
it should be such that one of the scalar fields oscillates around
its minimum, mimicking a pressureless dark-matter fluid, and the other field
induces a late-time accelerated expansion of the universe, thus behaving like
dark energy. Moreover, agreement with current cosmological observations restricts
the values of the parameters of the potential, namely, $V_a$ should be of the
order of $10^{-123}m_\texttt{P}^4$, and $\lambda$ should lie in the interval
$0\lesssim \lambda \lesssim 4$. In what concerns the parameter $\xi_a$,
the condition $m_\xi \gg H$ requires it to be much smaller than
$10^{-25} m_\texttt{P}$.

\section{Conclusions\label{Conclusions}}

In this article we have proposed a unified description of dark energy and dark
matter within the generalized hybrid metric-Palatini theory of gravity.

Working in the scalar-tensor representation of the theory we have shown that,
upon an appropriate choice of the potential [see Eq.~(\ref{potential DE-DM})],
one of the scalar fields behaves like dark energy, dominating the dynamics of
late-time cosmological evolution and yielding an accelerated expansion of the
universe, while the other scalar field behaves like pressureless dark matter
that, together with ordinary baryonic matter, dominates the intermediate
phases of cosmic evolution.

In order to ensure that the picture emerging from our unified description of
dark energy and dark matter is in agreement, at least qualitatively, with the
standard picture of the universe's evolution, we have imposed that, for each
and every solution of Eqs.~(\ref{Eq phi 3})--(\ref{Eq Dotdota 3}), the
present-time energy densities --- for radiation, ordinary baryonic matter,
dark matter, and dark energy --- should be equal to the measured ones, namely,
$\Omega_{\texttt{DE}0}=0.69$, $\Omega_{\texttt{DM}0}=0.26$,
$\Omega_{\texttt{BM}0}=0.05$, and $\Omega_{\texttt{R}0}=5\times10^{-5}$. This
requirement leaves us with just one free parameter in Eqs.~(\ref{Eq phi
3})--(\ref{Eq Dotdota 3}), $\lambda$, which we could, in principle, choose
freely. However, as our numerical simulations show, for $|\lambda-2|\gtrsim2$,
the transition from the radiation- to the matter-dominated era takes place so
early that it starts conflicting with primordial nucleosynthesis. Since this
is a conflict one wants to avoid, the parameter $\lambda$ was restricted to be
in the interval $0\lesssim\lambda\lesssim4$.

For such values of $\lambda$, our numerical solutions correspond to universes
in which the transition from a radiation- to the matter-dominated era occurs
well after primordial nucleosynthesis; the matter-dominated era is long enough
to allow, in principle, for structure formation to take place; the transition
to an era dominated by dark energy occurs in a recent past (see Figs.~\ref{Fig
Omega for lambda 2} and \ref{Fig Omega for lambda 0}). Furthermore, for
$\lambda\leq2\sqrt3$, the universe undergoes a late-time everlasting period of
accelerated expansion, during which the effective equation-of-state parameter
$w_{\rm eff}$ approaches the value $-1+\lambda^2/18$ (see Fig.~\ref{Fig w for
different lambda}). In short, the generalized hybrid metric-Palatini theory of
gravity provides a theoretical framework for a unified description of dark
matter and dark energy, giving rise to viable cosmological solutions, which
reproduce the standard phases of evolution of the universe.

In the present framework, it should be possible, in principle, to extend the
analysis backward in time in order to include an inflationary period driven
by the scalar field $\phi$ or $\xi$, thus achieving a triple unification of
inflation, dark matter, and dark energy within the generalized hybrid
metric-Palatini theory of gravity. Indeed, with our choice of the potential
$V(\phi,\xi)$ [see Eq.~(\ref{potential DE-DM})], any of the scalar fields
$\phi$ or $\xi$ could support an inflationary period; if, then, the decay of
this inflaton field were incomplete, the residue could play the role of dark
energy or dark matter in the manner proposed in the present article.

In fact, this idea has already been implemented in other contexts. For
example, the Salam-Sezgin six-dimensional supergravity theory yields, upon
compactification, an action containing two scalar fields $x$ and $y$ with
potential $V=A\exp(\sqrt2\kappa y)+ m_x^2 (x-x_{\rm min})^2/2 + \dots$, where
$A$ is a constant related to fundamental quantities of the theory and $m_x
\propto \exp(\sqrt2\kappa y/2)$ is the $y$-dependent mass of the scalar field
$x$; this potential, which is quite similar to ours, allows for a triple
unification of inflation, dark matter, and dark energy \cite{henriques-2009}.
As a second example, let us mention the string-landscape inspired model of
Ref.~\cite{liddle-2006,liddle-2008}, where a unified description of inflation,
dark matter, and dark energy was achieved with a single scalar field $\xi$
with  a potential $V=V_a+m_\xi^2\xi^2/2$, which is a particular case of ours
(for $\lambda=0$).

It would be interesting to implement this idea in the context of the
generalized hybrid metric-Palatini theory of gravity, allowing for the
unification of phenomena apparently as disparate as inflation, dark energy,
and dark matter within a single theoretical framework.

\begin{acknowledgments}
The author would like to thank Lucas S\'a for useful comments.
\end{acknowledgments}

\end{document}